# Realization of multifunctional shape-memory ferromagnets in all-*d*-metal Heusler phases

Z. Y. Wei,[1] E. K. Liu,[1,a)] J. H. Chen,[1] Y. Li,[1,2] G. D. Liu,[2] H. Z. Luo,[2] X. K. Xi,[1] H. W. Zhang,[1] W. H. Wang,[1] and G. H. Wu[1]

[1] *State Key Laboratory for Magnetism, Beijing National Laboratory for Condensed Matter Physics, Institute of Physics, Chinese Academy of Sciences, Beijing 100190, China*

[2] *School of Materials Science and Engineering, Hebei University of Technology, Tianjin 300130, China*

ABSTRACT

Heusler ferromagnetic shape-memory alloys (FSMAs) normally consist of transition-group *d*-metals and main-group *p*-elements. Here, we report the realization of FSMAs in Heusler phases that completely consist of *d* metals. By introducing the *d*-metal Ti into NiMn alloys, cubic B2-type Heusler phase is obtained and the martensitic transformation temperature is decreased efficiently. Strong ferromagnetism is established by further doping Co atoms into the B2-type antiferromagnetic Ni-Mn-Ti austenite. Based on the magnetic-field-induced martensitic transformations, collective multifunctional properties are observed in Ni(Co)-Mn-Ti alloys. The *d* metals not only facilitate the formation of B2-type Heusler phases, but also establish strong ferromagnetic coupling and offer the possibility to tune the martensitic transformation.

---

a) E-mail: ekliu@iphy.ac.cn

As important multifunctional materials, ferromagnetic shape-memory alloys (FSMAs) have been found in Heusler alloys,[1] Fe-based alloys[2] and hexagonal MM'X alloys.[3] All these alloys undergo a ferromagnetic martensitic transformation (FMMT) between different magnetic and structural states.[4,5] With a large magnetization difference (ΔM), the austenite and martensite phases gain Zeeman energy in different degrees in magnetic fields and the martensitic transformation (MT) temperature ($T_M$) can be tailored subsequently.[4] Numerous properties of FSMAs such as magnetic-field-induced shape-memory behavior,[4] magneto-strain,[6] magnetoresistance (MR),[7] exchange bias (EB),[8] magnetocaloric effect (MCE),[9] and electric power generation[10] have been extensively reported. Till now, studies on FSMAs have been most intensively conducted on Heusler alloys.[11,12]

In Heusler alloys, the main-group (*p*-group) atoms at the D sites form covalent bonds by *p-d* orbital hybridization with transition-metal (*d*-group) atoms at the (A,C) sites.[13,14] Firstly, the local atom configurations thus show high selectivity on chemical atoms according to the different types and strengths of orbital hybridizations.[15] Secondly, the electron pairing effect[16] from strong *p-d* covalent bonding considerably reduces the magnetic moments on the (A,C) sites, while the large moments at the B sites are maintained as the *p-d* covalent bonding becomes very weak between the next-nearest-neighbor B and D sites. Thirdly, the magnetic moments may form long-range order based on atomic ordering on the sites.[17] Last but very importantly, the strong bonding due to *p-d* hybridization may considerably stabilize the parent phase,[15] resulting in an extrinsic behavior that $T_M$ can be lowered with increasing content of *p*-group elements.[18,19] The *p-d* orbital hybridization in Heusler alloys thus plays a crucial role in the atom ordering, the magnetic structure, and the stabilization of the parent phase.

Apart from the *p-d* covalent hybridization, another important type of bonding, the covalent *d-d* hybridization between different *d*-group elements, also to some extend contributes to the formation and stabilization of Heusler phases. For example, the *d-d* hybridization between Ni and Mn in $Ni_2MnGa$ also influences the MT.[20] Recently, two-dimensional honeycomb hafnene has been reported[21] with significant covalent *d-d* bonding between Hf atoms which demonstrates that pure *d-d* bonding can stabilize the phase structure. More generally, one may expect that *d-d* hybridization can give rise to ordered structures, which is especially important for tuning of the phase transformations in FSMAs.

In the present study, we first introduce low-valence Ti into the binary NiMn alloy to create the $Ni_{50}Mn_{50-y}Ti_y$ and $Mn_{50}Ni_{50-y}Ti_y$ systems. In both cases, with increasing Ti content the austenite phase is stabilized and $T_M$ decreases remarkably. The substituted Ti promotes the formation of the Heusler phase by *d-d* covalent bonding of Ni/Mn and Ti atoms. By introducing Co atoms, strong ferromagnetism is established and FSMAs are realized in the $Ni_{50-x}Co_xMn_{50-y}Ti_y$ system, with various multifunctional properties.

$Ni_{50}Mn_{50-y}Ti_y$ ($y$ = 0, 6, 9, 12, 15, 20, 25 and 30; denoted as Ti$y$), $Mn_{50}Ni_{50-y}Ti_y$ ($y$ = 0, 6, 9, 10, 11, 12 and 15), and $Ni_{50-x}Co_xMn_{35}Ti_{15}$ ($x$ = 11.5, 12.5, 13.5 15, 15.9 and

17; denoted as Co$x$) alloys were prepared by arc melting high purity metals in argon atmosphere. The ingots were subsequently annealed in evacuated quartz tubes for six days at 1173 K for Ni$_{50}$Mn$_{50-y}$Ti$_y$ and Ni$_{50-x}$Co$_x$Mn$_{35}$Ti$_{15}$, and 1100 K for Mn$_{50}$Ni$_{50-y}$Ti$_y$ and then quenched in cold water. The magnetic properties were measured in a superconductive quantum interference device (SQUID) magnetometer. The martensitic and magnetic transition temperatures were determined by differential scanning calorimetry (DSC) and by magnetic measurements. Room temperature (RT) X-ray diffraction (XRD) was performed using Cu-$K\alpha$ radiation. The strain and electrical resistivity were measured with a physical property measurement system (PPMS). The morphology of martensite variants was probed by image quality mapping using the electron back-scattering diffraction (EBSD) technique.

Figure 1(a) shows the RT XRD patterns of Ni$_{50}$Mn$_{50-y}$Ti$_y$. For alloys with high Ti-contents (Ti20, Ti25, Ti30), a cubic austenite is found at RT, with, for example, a lattice parameter of $a$ = 5.9315 Å of the Ti25 alloy. The presence of the (200) superlattice reflection (Fig. 1(b)) indicates that the cubic phase has the ordered B2 structure. The (111) superlattice reflection would be indicative of the ordered L2$_1$-type structure of martensite.[14] The absence of this reflection in the XRD pattern may be attributed to atom occupancy disorder between the B and the D sites due to kinetic arrest.[22,23] Nevertheless, the very similar scattering factors of the 3$d$ metals will also prevent the occurrence of this superlattice reflection. In fact, the calculated relative intensity of this reflection is very low (I$_{(111)}$/I$_{(220)}$ ~ 0.3%) (Fig. 1(b)), so that it would be difficult to observe this possible reflection experimentally. In conventional Heusler alloys, it is known that the *d*-group atoms with more valence electrons preferably occupy (A,C) sites while *d*-group atoms with less valence electrons prefer the B sites and *p*-group atoms also with less valence electrons the D sites.[13,17] In present study, Ti atoms possess the lowest number of valence electrons ($3d^24s^2$) among the involved elements. When Ti is substituted for Mn atoms, the Ti atoms occupy the Mn(D) sites (Fig. 1(c1)), like *p*-group atoms. It is reasonable that the cubic austenite thus crystallizes in the B2-type (even $L2_1$-type) of structure with the (A,C) sites being occupied by Ni atoms, the B sites by Mn atoms and the D sites by Ti and residual Mn atoms (Fig. 1(c1)). Therefore, the Ni-Mn-Ti alloys composed of 3*d* metals can be structurally identified as Heusler phase with the B2-type (or $L2_1$-type) structure.

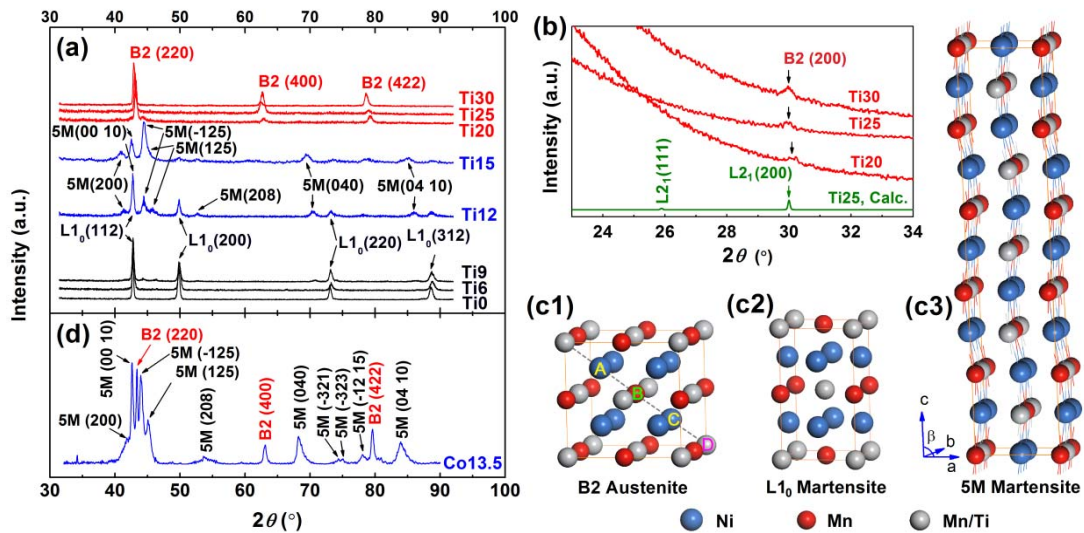

FIG. 1. Room-temperature XRD patterns and schematic crystal structures. (a) XRD patterns for $Ni_{50}Mn_{50-y}Ti_y$. (b) Step-scan XRD patterns of the superlattice reflections of the Ti20, Ti25 and Ti30 alloys. Schematic structures of (c1) B2 austenite, (c2) $L1_0$ martensite and (c3) 5-layer modulated (5M) martensite. (d) XRD patterns of the Co13.5 alloy.

In the XRD patterns of the alloys with low Ti contents (Ti0, Ti6, Ti9) (Fig. 1(a)), the $L1_0$ martensite structure is clearly observed. The lattice parameters of Ti6 are $a$ = 3.652 Å and $c$ = 7.317 Å, which is consistent with the $L1_0$ structure reported for the alloy NiMn.[24] In the alloys Ti12 and Ti15 with medium Ti contents, coexistence of different martensites with $L1_0$ and 5-layer modulated (5M) structures is found. The 5M structure is a monoclinic system (space group P2/m). The structural parameters of Ti15 are $a$ = 4.410 Å, $b$ = 5.407 Å, $c$ = 21.272 Å, and $\beta$ = 90.95°. The 5M martensite structure is schematically presented in Fig. 1(c3). Upon substitution of Co atoms at (A,C)(Ni) sites in $Ni_{50}Mn_{35}Ti_{15}$, the alloys remain a similar 5M structure, as indicated by XRD patterns of Co13.5 alloy (Fig. 1(d)). A large volume shrink of 1.95 % is derived from the parent and martensite phases during the MT around room temperature.

Figure 2 shows the temperature dependence of the magnetization in a magnetic field of 100 Oe for all studied alloys. For Ti20 (Fig. 2(a)) clear thermal hysteresis is observed between 181 and 194 K, which implies a first-order MT below RT. The MT above RT of Ti15 has been detected by DSC (left inset of Fig. 2(a)), showing exothermic and endothermic peaks at 395 and 415 K, respectively. With increasing Ti content, the austenite is stabilized completely and the MT disappears, leaving an antiferromagnetic (AFM)-like ordering ($T_N$) at low temperatures (120 K for Ti25 and 107 K for Ti30). For Ti20, a linear fit of $M^{-1}$ *vs* $T$ above the MT, with a negative extrapolated temperature, confirms the AFM character of the austenite. The linear magnetic isotherms at 200 K (austenite state) and 150 K (martensite state) (right inset of Fig. 2(a)) indicate paramagnetic (PM) behavior of the austenite and PM (or AFM) behavior of the martensite. It is seen that by introducing Ti, the MT of NiMn can be tuned to low temperatures by stabilizing the austenite. However, the introduced Ti cannot establish strong ferromagnetism in the Ni-Mn-Ti $d$-metal Heusler alloys.

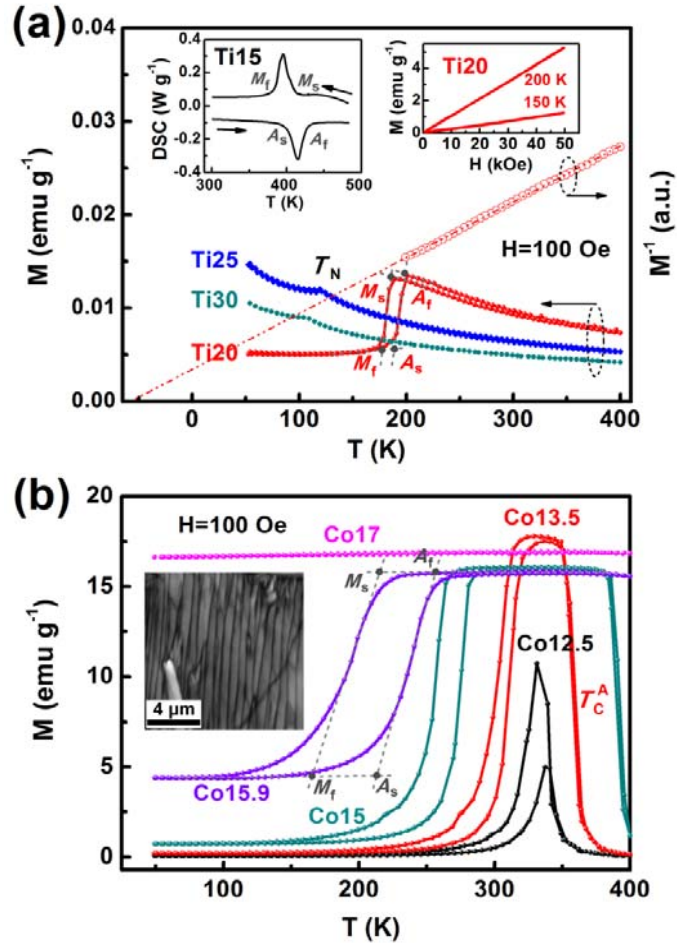

FIG. 2. (a) Thermomagnetic curves for $Ni_{50}Mn_{50-y}Ti_y$ (Ti20, 25 and 30) and $M^{-1}$ vs $T$ for Ti20. The left inset shows the DSC result for $Ni_{50}Mn_{35}Ti_{15}$. The right inset shows the magnetic isotherms of Ti20 well above and below the MT. $M_s$ and $M_f$ refer, respectively, to the starting and finishing temperatures of forward MT. $A_s$ and $A_f$ refer, respectively, to the starting and finishing temperatures of inverse MT. (b) Thermomagnetic curves of $Ni_{50-x}Co_xMn_{35}Ti_{15}$. The inset shows the morphology of 5M martensite variants of Co13.5 at room temperature.

The ferromagnetism can be introduced by substituting Co into the Ni-Mn-Ti alloys. The *M-T* curves of $Ni_{50-x}Co_xMn_{35}Ti_{15}$ are presented in Fig. 2(b) where abrupt magnetic transitions, corresponding to MTs from strong FM austenite to weak-magnetic martensite can be seen. Compared with the $Ni_{50}Mn_{35}Ti_{15}$ starting alloy, Co substitution results not only in a decrease of $T_M$, but also to the desired increase of the Curie temperature ($T_C$) of the austenite. It is found that in the $Ni_{50-x}Co_xMn_{35}Ti_{15}$ system, $T_M$ and $T_C$ merge at about 340 K in the Co12.5 alloy. For $x > 15.9$, the MT disappears and $T_C$ becomes higher than 400 K. It shows that, in Co-substituted Ni-Mn-Ti Heusler alloys, FMMTs can be realized. The image quality mapping of Co13.5 alloy, as shown in the inset to Fig. 2(b), show the fine morphology of plate-like 5M martensite variants with a plate thickness of about 500 nm.

Based on the XRD, DSC and magnetic measurements, structural and magnetic phase diagrams of the systems $Ni_{50}Mn_{50-y}Ti_y$, $Mn_{50}Ni_{50-y}Ti_y$ and $Ni_{50-x}Co_xMn_{35}Ti_{15}$ are proposed as depicted in Fig. 3. In Fig. 3(a) can be seen that, upon substitution of Ti for Mn in $Ni_{50}Mn_{50-y}Ti_y$, $T_M$ decreases very strongly. This behavior is consistent with conventional Heusler alloys, which are stabilized by *p-d* orbital hybridization of *p*-block (Al, Ga, In, Sn, Sb) elements.[18,19] In the present study, the Ti atoms occupy the Mn(D) sites and play a similar role in stabilizing the austenite phase by *d-d* orbital *d-d* hybridization with nearest-neighbor Ni atoms. Nevertheless, rather different from many conventional Heusler alloys (for example, $Ni_2MnGa$), the austenite of Ni-Mn-Ti Heusler alloys is an antiferromagnet with a low $T_N$. This may be related to the AFM exchange interaction controlled by both distance between magnetic moments and number of itinerant electrons of the system, described by the Stearns model.[25] Similar results of a decreasing $T_M$ and weak magnetism are obtained for the $Mn_{50}Ni_{50-y}Ti_y$ system (Fig. 3(a)).

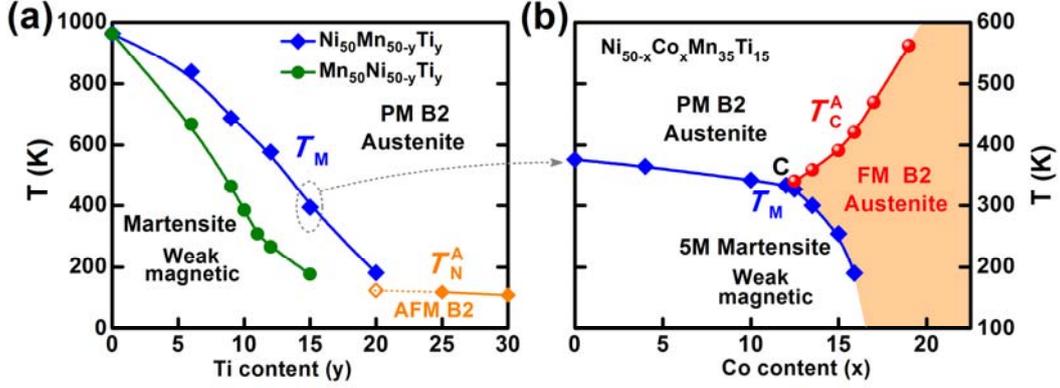

FIG. 3. Structural and magnetic phase diagrams of (a) $Ni_{50}Mn_{50-y}Ti_y$ and $Mn_{50}Ni_{50-y}Ti_y$ and (b) $Ni_{50-x}Co_xMn_{35}Ti_{15}$. $T_M = (M_s + M_f)/2$ (see caption of Fig. 2).

Figure 3(b) shows that, besides further decreasing $T_M$, Co substitution efficiently increases $T_C$. After the crossover point C, magnetostructural coupling begins to appear. In Co-doped Mn-rich Ni(Co)-Mn-X (X=Al, Ga, Sn, Sb) Heusler alloys, the ferromagnetic activation effect has been applied to increase the ferromagnetism of austenite by forming the Mn(B)-Co(A,C)-Mn(D) configuration,[26-33] in which the Co atoms occupy Ni(A,C) sites since both Ni and Co have more valence electrons than Mn and the X elements.[13,14] In the present B2-type $Ni_{50-x}Co_xMn_{35}Ti_{15}$ alloys, a similar situation exists. Co atoms that have been substituted for Ni atoms will also share the (A,C) sites with Ni atoms, leaving Mn and Ti with less valence electrons at the B/D sites. With the aid of the strong FM exchange interactions between nearest-neighbor Co-Mn atoms, the original AFM exchange coupling between Mn-Mn atoms in Ni-Mn-Ti alloys is converted into FM one, resulting in parallel alignment of the Mn-Co-Mn moments. The present experimentally observed strong ferromagnetism provides direct evidence of this probable atomic configuration and of the ferromagnetic activation effect in the Ni(Co)-Mn-Ti all-*d*-metal Heusler alloys. The detailed magnetic-exchange mechanism in these 3*d*-metal Heusler alloys needs further investigation.

Figure 4(a) shows the temperature dependence of the magnetization of the Co15 alloy in magnetic fields of 0.1 and 120 kOe. The field of 120 kOe shifts the MT to low temperatures by more than 20 K, as the magnetic field stabilizes the FM high-temperature phase. A large magnetization difference ($\Delta M$) of about 90 emu g$^{-1}$ is associated with the transformation, which will facilitate transformation-related magnetoresponsive effects. Magnetic isotherms, measured across the MT, are shown in Fig. 4(b). At 290 K, the magnetization reveals typical FM behavior of the austenite. The isotherms at lower temperatures of 235, 253 and 270 K exhibit a clear metamagnetic behavior with magnetic hysteresis, associated with the magnetic-field-induced reverse MTs from weak-magnetic martensite to strong-ferromagnetic austenite. Especially, at 253 K a two-way reversible MT is observed. At low temperatures (in particular at 5 K), the saturation magnetization and saturation field are both low (0.47 $\mu_B$ f.u.$^{-1}$ and 1.3 kOe, respectively), which implies that the martensite is magnetically ordered, with a possibility of ferrimagnetic structure, similar to NiCoMnSb[27] and Mn-based[34] Heusler compounds. This low magnetization of the martensite benefits the large $\Delta M$ of Ni(Co)-Mn-Ti alloys.

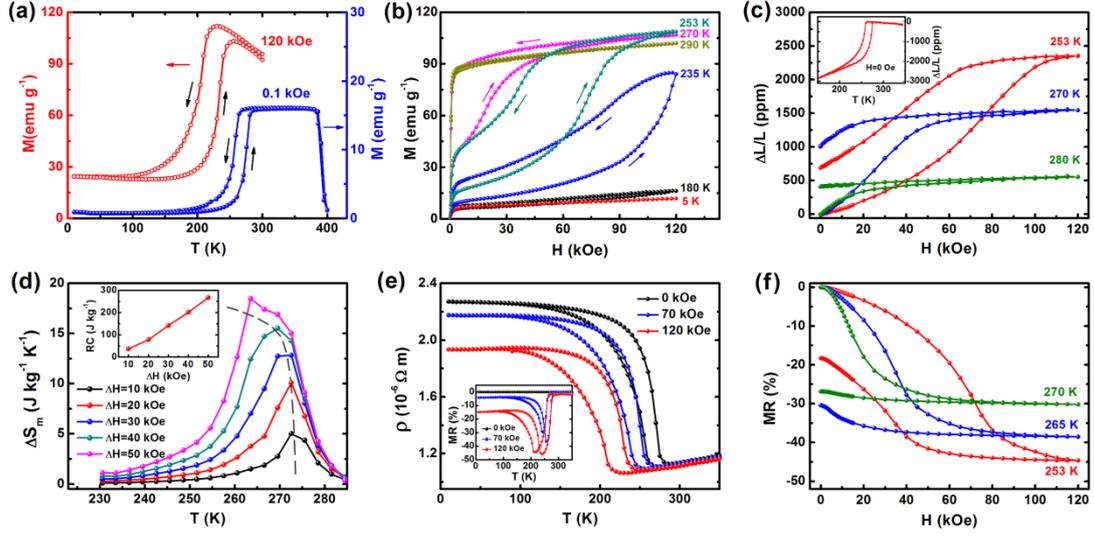

FIG. 4. Magnetic and magnetoresponsive properties across the MT of the Co15 alloy. (a) Temperature dependence of the magnetization in magnetic fields of 0.1 and 120 kOe. (b) Magnetic isotherms at several temperatures. (c) Field-induced strain at several temperatures. The inset shows the temperature dependence of the strain. (d) Magnetic-entropy change ($\Delta S_m$) at various field changes. The inset shows the refrigerating capacity (RC). (e) Temperature dependence of the electrical resistivity in magnetic fields of 0, 70, and 120 kOe. The inset shows the temperature dependence of the magnetoresistance (MR). (f) Field dependence of the MR at various temperatures.

Having obtained magnetic-field-induced FMMTs, we have further investigated a number of physical effects across the MTs. Figure 4(c) shows the field dependence of the strain of the Co15 alloy. At 253 K, the strain saturates with a maximum value of 2400 ppm in a magnetic field of 120 kOe. This large polycrystalline magneto-strain originates from cell-volume expansion due to the field-induced reverse MT. The strain recovery during decrease of the field indicates magnetic superelasticity of the alloys.[35] The inset of Fig. 4(c) shows the temperature dependence of the strain across the MT in zero field, with a transformation-related strain of about 2500 ppm. Further increase of the temperature results in a smaller saturation strain but also to a reduced saturation field. At 270 K, a magnetic field of 50 kOe can induce a saturation strain of 1300 ppm. Around room temperature (280 K), the strain tends to saturate at 300 ppm in a moderate field of 20 kOe, which will benefit the magnetic-field-induced shape-memory behavior.

The field-induced metamagnetic behavior has prompted us to study the magnetocaloric effect (MCE) in these $d$-metal FSMAs. Adopting the temperature-loop process method,[36] a series of magnetic isotherms with a temperature interval of 3 K was measured across the FMMT. Based on these measurements, the magnetic entropy changes ($\Delta S_m$) were calculated using Maxwell relation,[36] as shown in Fig. 4(d). At about 263 K, a maximum $\Delta S_m$ value around 18 J kg$^{-1}$ K$^{-1}$ is obtained for a field change of 50 kOe. This value of $\Delta S_m$ is comparable to the values for many other Heusler alloys. The inset of Fig. 4(d) shows the refrigerant capacity (RC), which has a relatively high value of 267 J kg$^{-1}$ for a field change of 50 kOe, showing that all-$d$-metal Heusler FSMAs may have a potential as magnetocaloric materials.

The MT is always accompanied by an electrical-resistance change. To investigate this, the temperature dependence of the electrical resistance was measured in various magnetic fields, as can be seen in Fig. 4(e), the electrical resistance exhibits a jump across the FMMT which is due to intrinsic changes of both the anisotropy and the area of the Fermi surface caused by the lattice distortion, which also has been observed for NiMnIn[7] and Ni$_2$FeGa[37] Heusler alloys. A remarkable shift of the jump onset with increasing field towards lower temperatures is related to the field-induced

FMMT. As illustrated in the inset of Fig. 4(e), quite large MR values of 37% at 70 kOe and 46% at 120 kOe are found across the transformation. The field dependence of the MR at different temperatures is shown in Fig. 4(f), where a field dependence of the MR is observed which is consistent with the field dependence of the strain. Based on the FMMTs, one can see different resistance states can be induced by both temperature and magnetic field in these *d*-metal Heusler alloys.

To conclude, a kind of promising shape-memory ferromagnets have been obtained in Heusler phases which only consist of 3d elements. Introduction of *d*-metal Ti into NiMn not only facilitates the formation of B2-type (even $L2_1$-type) Heusler phases, but also stabilizes the parent phase and tunes the MT. Starting from the B2-type Ni-Mn-Ti Heusler phase, Co substitution gives rise to strong ferromagnetic coupling in the antiferromagnetic austenite of Ni(Co)-Mn-Ti alloys. In these alloys, MTs with large magnetization difference can be driven by an external field, offering interesting functional properties. Our present study from the view of *d-d* hybridization deepens the understanding on the phase formation, martensitic transformation, and magnetic coupling of multifunctional shape-memory ferromagnets.

ACKNOWLEDGEMENTS
This work was supported by National Natural Science Foundation of China (51301195, 11174352, 51371190 and 51431009), National Basic Research Program of China (2012CB619405), Beijing Municipal Science & Technology Commission (Z141100004214004), and Youth Innovation Promotion Association of Chinese Academy of Sciences (2013002).